\documentclass{article}
\usepackage{amsfonts}
\usepackage{amssymb}
\usepackage{amsmath}
\usepackage[dvips]{graphics}
\usepackage{latexsym}
\usepackage{subfigure}

\newcommand{\be}{\begin{eqnarray}}
\newcommand{\ee}{\end{eqnarray}}
\newcommand{\bea}{\setlength\arraycolsep{2pt} \begin{eqnarray}}
\newcommand{\eea}{\end{eqnarray}}
\newcommand{\nn}{\nonumber}

\newcommand{\bean}{\begin{eqnarray}}
\newcommand{\eean}{\end{eqnarray}}
\newcommand{\eqs}[1]{Eqs. (\ref{#1})}
\newcommand{\eq}[1]{Eq. (\ref{#1})}
\newcommand{\meq}[1]{(\ref{#1})}

\newcommand{\ppa}[2]{\left(\frac{\partial}{\partial #1}\right)^{#2}}

\newcommand{\eqn}{&=&}
\newcommand{\non}{\nonumber \\}

\newcommand{\te}[1]{{(#1)}}

\def\0{{\sst{(0)}}}
\def\1{{\sst{(1)}}}
\def\2{{\sst{(2)}}}
\def\3{{\sst{(3)}}}
\def\4{{\sst{(4)}}}
\def\5{{\sst{(5)}}}
\def\6{{\sst{(6)}}}
\def\7{{\sst{(7)}}}
\def\8{{\sst{(8)}}}
\def\sst#1{{\scriptscriptstyle #1}}

\title{Kerr black holes as accelerators of spinning test particles}
\author{ Minyong Guo\thanks{Email: minyongguo@mail.bnu.edu.cn} and Sijie Gao\thanks{Corresponding author. Email: sijie@bnu.edu.cn} \\
Department of Physics, Beijing Normal University,\\
Beijing 100875, China}

\begin{document}
\maketitle

\begin{abstract}
It has been shown that ultraenergetic collisions can occur near the horizon of an extremal Kerr black hole. Previous studies mainly focused on geodesic motions of particles. In this paper, we consider spinning test particles whose orbits are non-geodesic. By employing the  Mathisson-Papapetrou-Dixon equation, we find the critical angular momentum satisfies $J=2E$ for extremal Kerr black holes. Although the conserved angular momentum $J$ and energy $E$ have been redefined in the presence of spin, the critical condition remains the same form. If a particle with this angular momentum collides with another particle arbitrarily close to the horizon of the black hole, the center-of-mass energy can be arbitrarily high. We also prove that arbitrarily high energies cannot be obtained for spinning particles near the horizons of non-extremal Kerr black holes.\\

PACS number(s):   04.70.Bw, 97.60.Lf

\end{abstract}

\section{Introduction}
In 2009, Ba\~nados, Silk and West \cite{bsw} showed that an extremal black hole can be used as a particle accelerator and arbitrarily high energy could be produced near the event horizon. Since their pioneering work, this issue has been extensively studied for various black holes \cite{vitor}-\cite{zas}. Although the idea of infinite energy is fascinating, there are practical limitations on the BSW mechanism. Since astrophysical black holes have the maximum spin-mass ratio  $a/M=0.998$, it follows that the maximum center-of-mass energies by the BSW mechanism is only about 10 times the rest mass of the particle \cite{vitor}. Moreover, it was shown explicitly \cite{prl2012} that even if the energy of collision diverges near the horizon, only a limited amount of energy (approximately the rest mass of the colliding particles) can reach a distant observer.

 So far, most authors focus on studying geodesic motions of point particles. A real particle is an extended body with self-interaction. It has been shown \cite{math}-\cite{wald} that the motion of a spinning test particle deviates from a geodesic due to gravitational interaction. The orbits of spinning particles around black holes have been calculated based on the Mathisson-Papapetrou-Dixon(MPD) equation \cite{prd98}-\cite{inner}.

The essential part of the BSW mechanism is that a particle with a critical angular momentum will approach the black hole horizon asymptotically. Since the motion of a spinning particle is no longer a geodesic, it is unclear whether the BSW mechanism works. We shall address this issue in this paper. For an extremal Kerr black hole, we calculate the conserved quantities derived from the MPD equation and find the critical angular momentum that could produce infinite center-of-mass energy at the horizon. It is interesting to see that the critical angular momentum-energy relation has the same form as that for non-spinning particles, although both the angular momentum and  energy have been redefined in the presence of spin.

An important result of the BSW mechanism is that infinite energy is not possible if the black hole is non-extremal\cite{jacobson,Gao}. This property significantly restricts the application of the BSW mechanism because  an astrophysical black hole can not be extremal, as we mentioned at the beginning of this section.
By applying the MPD equation again, a more complicated calculation shows that a non-extremal black hole can not be used as an accelerator to produce arbitrarily large energy even for particles with spin.

It is worth mentioning that the collision of spinning particles near Schwarzschild black holes was studied in \cite{ar15}. We shall further discuss  the result of \cite{ar15} in section \ref{coll}.

\section{Equation of motion of a spinning test body }
To describe the influence of body's spin on its orbits, we employ the Mathisson-Papapetrou-Dixon(MPD) equations \cite{wald}:
\be
  \frac{Dp^a}{D\tau}=-\frac{1}{2}R^a_{~bcd}v^bS^{cd} \,,\nn
  \ee
  \be
  \frac{D S^{ab}}{D\tau}=p^av^b-p^bv^a\,,
  \ee
where
\bean
v^a=\ppa{\tau}{a}   \label{tva}
\eean
is the tangent to the center-of-mass world line,
$p^a$ is the 4-momentum of the body, $\frac{D}{D\tau}$ is the covariant derivative along the world line, and $S^{ab}$ is its spin tensor. The quantity
\bean
m^2\equiv-p^ap_a
\eean
is conserved along the orbit, and we regard it as the square of the mass of the particle. We also introduce the dynamical velocity,
\bean
u^a\equiv \frac{p^a}{m}\,,
\eean
which is in general not tangent to the world line of the spinning particle.
The equations were derived under the assumption that characteristic radius of the spinning particle is much smaller than the curvature scale of a background spacetime and the mass of a spinning body is much less than that of black holes.
   In order to close the system, we impose the supplementary condition\cite{prd98}
   \be
   S^{ab}p_b=0\,.
 \ee
 This relation fixes the center-of-mass of the spinning particle. Since the magnitude of spin is also conserved,  we set
    \be
    S^{ab}S_{ab}=2S^2=2m^2s^2\,,
    \ee
    where the constant $s=S/m$ is interpreted as the specific spin angular momentum of the particle. For latter convenience, the parameter $\tau$ in \eq{tva} is normalized as
 \be
 u^av_a=-1\,.
 \ee
So $\tau$ is not the proper time of the particle.

The above equations also imply the relation between $u^a$ and $v^a$ \cite{prd98}:
    \be
    v^a-u^a=\frac{S^{ab}R_{bcde}u^cS^{de}}{2(m^2+\frac{1}{4}R_{bcde}S^{bc}S^{de})}\,. \label{vur}
    \ee

If  $\xi^a$ is a Killing vector field which satisfies $\nabla_{(a}\xi_{b)}=0$ , one can show that the quantity
      \be
      Q_\xi=p^a\xi_a-\frac{1}{2}S^{ab}\nabla_b\xi_a \label{cons}
      \ee
is constant along the particle's trajectory. The conserved quantities will be very useful to find the orbits of the spinning particle.

\section{Spinning body in Kerr spacetime}
\subsection{Tetrad bases and conserved quantities}
For the Kerr solution \cite{waldbook}
\bean
ds^2\eqn -\left(\frac{\Delta-a^2\sin^2\theta}{\Sigma}\right)dt^2-
\frac{2a\sin^2\theta(r^2+a^2-\Delta)}{\Sigma}dt d\phi\non
&+&\left[\frac{(r^2+a^2)^2-\Delta a^2\sin^2\theta}{\Sigma}\right]\sin^2\theta d\phi^2+\frac{\Sigma}{\Delta} dr^2+\Sigma d\theta^2\,,
\eean
where
\bean
\Sigma\eqn r^2+a^2\cos^2\theta\,, \\
\Delta\eqn r^2+a^2-2Mr \,,
\eean
the tetrad reads \cite{prd98}
\bean
e^\te{0}_a\eqn\sqrt{\frac{\Delta}{\Sigma}}(dt_a-a\sin^2\theta d\phi)\,, \label{et1} \\
e^\te{1}_a\eqn\sqrt{\frac{\Sigma}{\Delta}}dr_a \label{et2}\,, \\
e^{(2)}_a\eqn \sqrt{\Sigma} d\theta_a \,, \label{et3}  \\
e^{(3)}_a\eqn\frac{\sin\theta}{\sqrt\Sigma}[-adt_a+(r^2+a^2)d\phi_a] \,. \label{et4}
\eean

There are two Killing vectors: the timelike Killing vector $\xi^a$,
\be
\xi^a=\ppa{t}{a}\,,
\ee
and the axial Killing vector $\phi^a$,
\be
\phi^a=\ppa{\phi}{a}\,.
\ee
They correspond to two  conserved quantities: the energy $\tilde{E}$, and the $z$ component of the total angular momentum $\tilde{J}$.  By applying \eq{cons}, we have
\bea
\frac{\tilde{E}}{m}&=&-u^a\xi_a+\frac{1}{2m}S^{ab}\nabla_b\xi_a \,, \non
\frac{\tilde{J}}{m}&=&u^a\phi_a-\frac{1}{2m}S^{ab}\nabla_b\phi_a \,.\label{ejm}
\eea

By calculating the tetrad components of \eq{ejm}, $\tilde{E}$ and $\tilde{J}$ are given by \cite{prd98}
\bean\label{5}
\frac{\tilde{E}}{m}
&=&\sqrt{\frac{\Delta}{\Sigma}}u^{(0)}+\frac{a\sin\theta}{\sqrt{\Sigma}}u^{(3)}
+\frac{M(r^2-a^2\cos^2\theta)}{\Sigma^2}\frac{S^{(1)(0)}}{m}+
\frac{2Mar\cos\theta}{\Sigma^2}\frac{S^{(2)(3)}}{m} \,,\label{em} \non
\frac{\tilde{J}}{m}
&=&a\sin^2\theta\sqrt{\frac{\Delta}{\Sigma}}u^{(0)}+\frac{(r^2+a^2)
\sin\theta}{\sqrt{\Sigma}}u^{(3)}\non
&&+\frac{a\sin^2\theta}{\Sigma^2}[(r-M)\Sigma+2Mr^2]\frac{S^{(1)(0)}}{m}+
\frac{a\sqrt{\Delta}\sin\theta\cos\theta}{\Sigma}\frac{S^{(2)(0)}}{m}\non
&&+\frac{\cos\theta}{\Sigma^2}[(r^2+a^2)^2-a^2\Delta\sin^2\theta]\frac{S^{(2)(3)}}{m}
+\frac{r\sqrt{\Delta}\sin\theta}{m\Sigma}S^{(1)(3)} \,. \label{jm}
\eean

\subsection{Equations of motion on the equatorial plane}
In the following, we shall consider the case where a spinning particle moves on the equatorial plane ($\theta=\frac{\pi}{2}$) of the Kerr spacetime.  We first introduce a specific spin vector $s^{(a)}$ as
\be
s^{(a)}=-\frac{1}{2m}\varepsilon^{(a)}_{~~~(b)(c)(d)}u^{(b)}S^{(c)(d)} \,,
\ee
or equivalently
\be\label{12}
S^{(a)(b)}=m\varepsilon^{(a)(b)}_{~~~~~~(c)(d)}u^{(c)}s^{(d)} \,,
\ee
where $\varepsilon_{(a)(b)(c)(d)}$ is the completely antisymmetric tensor with the component $\varepsilon_{(1)(2)(3)(4)}=1$. By the argument in \cite{prd98}, one may set the only non-vanishing component of $s^\te{a}$ to be
\bean
s^{(2)}=-s  \,,
\eean
where $s$ indicates not only the magnitude of spin but also the spin direction. The particle's spin is parallel to the black hole spin for $s>0$, while it is antiparallel for $s<0$.
Consequently, the nonvanishing tetrad components of the spin angular momentum are given by
\bean
S^{(0)(1)}&=&-msu^{(3)} \,, \non
S^{(0)(3)}&=&msu^{(1)} \,, \non
S^{(1)(3)}&=&msu^{(0)} \,. \label{bis}
\eean
We define $E\equiv\frac{\tilde{E}}{m}$ and $ J\equiv\frac{\tilde{J}}{m}$ as the energy per unit mass, and the angular momentum per unit  mass. By substituting \eq{bis} into  \meq{jm}, we find at the equatorial plane $\theta=\frac{\pi}{2}$:
\bean\label{con}
E&=&\frac{\sqrt{\Delta}}{r}u^{(0)}+\frac{ar+Ms}{r^2}u^{(3)} \,,\label{ee}\\
J&=&\frac{\sqrt{\Delta}}{r}(a+s)u^{(0)}+[\frac{r^2+a^2}{r}+\frac{as}{r^2}(r+M)]u^{(3)}\,. \label{jj}
\eean
By calculating \eq{vur},
the relation between the normalized momentum vector $u^{(a)}$ and the 4-velocity $v^{(a)}$ can be expressed as \cite{prd98}
\bean
v^{(0)}&=&N(1-\frac{Ms^2}{r^3})u^{(0)}\,,\non
v^{(1)}&=&N(1-\frac{Ms^2}{r^3})u^{(1)}\,, \non
v^{(3)}&=&N(1+\frac{2Ms^2}{r^3})u^{(3)}\,,\label{relation}
\eean
where
\be
N=\left(1-\frac{Ms^2}{r^3}[1+3(u^{(3)})^2]\right)^{-1} \,.
\ee
\section{Collisions near the event horizon $r=r_+$} \label{coll}
In this section,we consider two spinning particles  colliding outside a Kerr black hole. We shall investigate whether arbitrarily high center-of-mass energies can be obtained.

The 4-vector of a spinning particle at the point of collision takes the general form
\be
v^a=\frac{dt}{d\tau}\ppa{t}{a}+\frac{dr}{d\tau}\ppa{r}{a}+\frac{d\phi}{d\tau}\ppa{\phi}{a}\,.
\ee
Using \eqs{et1}-\meq{et4} , we can get
\be\label{7}
\nn
v^{(0)}&=&\sqrt{\frac{\Delta}{\Sigma}}\left(\frac{dt}{d\tau}-a\sin^2{\theta}
\frac{d\phi}{d\tau}\right) \,, \\\nn
v^{(1)}&=&\sqrt{\frac{\Sigma}{\Delta}}\frac{dr}{d\tau}\,,\\
v^{(3)}&=&\frac{\sin{\theta}}{\sqrt{\Sigma}}\left[-a\frac{dt}{d\tau}+
(r^2+a^2)\frac{d\phi}{d\tau}\right]  \,.
\ee
With the help of \eq{ee} and \eq{jj}, we obtain
\be\label{8}
u^{(0)}&=&\frac{r \left(a^2 E r+a (E s (M+r)-J r)+E r^3-J M s\right)}{\sqrt{\Delta} \left(r^3-M s^2\right)}\,,\\
u^{(3)}&=&-\frac{r^2 (a E+E s-J)}{r^3-M s^2} \,.
\ee
Then the normalization condition $-(u^\te{0})^2+(u^\te{1})^2+(u^\te{3})^2 =-1$ yields
\bean
u^\te{1}=-\frac{1}{\sqrt{\Delta}(r^3-Ms^2)}\sqrt{O}\,, \label{uone}
\eean
where
\be
O&=&a^4E^2r^4+(Er^4-JMrs)^2+2a^3Er^3[-Jr+E(M+r)s]\non
&&-\Delta(J^2r^4+r^6-2EJr^4s-2Mr^3s^2+E^2r^4s^2+M^2s^4)\non
&&-2ar^2[-J^2Mrs-E^2r^3(M+r)s+\Delta Er^2(Es-J)+EJ(r^4+M^2s^2+Mrs^2)\non
&&+a^2r^2[-\Delta E^2r^2+J^2r^2-2EJr(2M+r)s+E^2(2r^4+M^2s^2+2Mrs^2+r^2s^2)] \,.\non
\ \  \label{bigo}
\ee
Note that we have chosen the minus sign for $u^{(1)}$ because we are interested in  ingoing orbits. Using Eqs. (\ref{relation}), (\ref{7})-(\ref{bigo}), the explicit components of velocity fields can be derived as \cite{prd98, inner}
\be\label{14}
\nn
\Sigma_s\Lambda_s\frac{dt}{d\tau}&=&a(1+\frac{3Ms^2}{r\Sigma_s})[J-(a+s)E]+\frac{r^2+a^2}{\Delta}P_s\,,\\\nn
\Sigma_s\Lambda_s\frac{dr}{d\tau}&=&-\sqrt{R_s}\,,\\
\Sigma_s\Lambda_s\frac{d\phi}{d\tau}&=&(1+\frac{3Ms^2}{r\Sigma_s})[J-(a+s)E]+\frac{a}{\Delta}P_s \,,
\ee
where
\be
\nn
\Sigma_s&=&r^2(1-\frac{Ms^2}{r^3})\,,\\\nn
\Lambda_s&=&1-\frac{3Ms^2r[-(a+s)E+J]^2}{\Sigma_s^3}\,,\\\nn
R_s&=&P_s^2-\Delta(\frac{\Sigma_s^2}{r^2}+[-(a+s)E+J]^2)\,,\\
P_s&=&[(r^2+a^2)+\frac{as}{r}(r+M)]E-(a+\frac{Ms}{r})J  \,.
\ee

We can also have
\be\label{9}
v^av_a=-\frac{P(r^3-Ms^2)^2}{(r^9-3Mr^6s^2+3M^2r^3s^4-M^3s^6-3Mr^4s^2[J-E(a+s)]^2)^2} \,,
\ee
where
\be\label{pp}
\nn
P&=&(r^{12}-4Mr^9s^2+6M^2r^6s^4-4M^3r^3s^6+M^4s^7\\
&-&6Mr^7s^2[J-E(a+s)]^2-3M^2r^4s^4[J-E(a+s)]^2) \,.
\ee
Since  $v^av_a$ is the tangent to the timelike wordline, it follows  that $v^av_a<0$, which means
\bean
P>0  \label{pb} \,.
\eean
We shall check this condition later for  relevant orbits.

Obviously, a physically allowed trajectory satisfies
\be\label{17}
O\ge0  \,.
\ee
Suppose two spinning particles with the same mass $m$ collide into each other. The center-of-mass energy is given by
\be
E_{c.m.}=\sqrt{2}m\sqrt{1-g_{ab}n^an^b_1} \,,
\ee
where $n_a=\lambda v_a$ satisfying $n_an^a=-1$, and $n_1^a$ is the normalized tangent for the other particle.

For simplicity, we define the effective center-of-mass energy \cite{Gao}
\be\label{15}
E_{eff}=-g_{ab}n^an^b_1  \,.
\ee
Setting $k=\lambda N(1-\frac{Ms^2}{r^3})>0, l=\lambda N(1+\frac{2Ms^2}{r^3})>0$,
we find
\be
E_{eff}=ll_2\frac{r^4(a E-J+Es)(a E_1-J_1+E_1s_1)}{(r^3-Ms^2)(r^3-Ms_1^2)}-k k_1\frac{E^{\prime}}{\Delta(r^3-Ms^2)(r^3-Ms_1^2)}  \,,\non
\label{eef}
\ee
where
\be\label{ep19}
\nn
E^{\prime}&=&-r^2[a^2Er+Er^3-JMs+a(-Jr+EMs+Ers)][a^2E_1r+E_1r^3-J_1Ms_1\\
&+&a(-J_1r+E_1Ms_1+E_1rs_1)]+\sqrt{O}\sqrt{O_1} \,,
\ee
where $O$ is given by \eq{bigo} and $O_1$ is obtained by replacing $J$,$E$ and $s$ with $J_1$,$E_1$ and $s_1$ in \eq{bigo}.
Without loss of generality, we shall choose
\be
M=1 \,.
\ee
Since $n^a$ is a future-directed timelike vector and $\lambda>0$, it follows that $\frac{dt}{d\tau}>0$ near the horizon $r=r_+$.We rewrite Eq. (\ref{14}) as
\bean
\frac{dt}{d\tau}=\frac{K}{[r^9-3r^6s^2+3r^3 s^4-s^6-3r^4s^2(J-Eq-Es)^2]\Delta}\,.
\eean
The expression for $K$ is lengthy. Since we shall be interested in the sign of $dt/d\tau$ near the horizon, we find
\bean
K\sim  (r^3-Ms^2)(r^2+a^2)(r^2-s^2)[-J(ar+s)+E(a^2 r+r^3+as+ars)]\,.
\eean
as $r\rightarrow r_+$. Because $s$ is small, the condition $\frac{dt}{d\tau}>0$ implies
\bean
-J(ar+s)+E(a^2 r+r^3+as+ars)>0  \label{jin}
\eean
for $r\rightarrow r_+$. Our purpose is to examine whether an infinite $E_{eff}$ defined in Eq.(\ref{15}) exists under the constraints \meq{pb}, (\ref{17}) and (\ref{jin}).

It is easy to see from \eq{eef} that an infinite $E_{eff}$ cannot be obtained unless one of the denominators vanishes. The vanishing of the first denominator indicates a radial turning point $r_s$ satisfying
\bean
r_s=M\left(\frac{s}{M}\right)^{2/3} \,.\label{rs}
\eean
The same expression was found in \cite{ar15} in a Schwarzschild background. However, as analyzed in \cite{ar15}, such a divergence of energy may not be real because superluminal motions must occur. Here, we provide a  simpler argument to rule out this possibility. As shown by M$\o$ller \cite{moller} (see also \cite{wald} for a simple explanation), the positive energy density of a body implies that the size $r_0$, the spin $S$, and the mass $m$ of the body must satisfy
\bean
r_0\gtrsim S/m \,.
\eean
So we have
\bean
s=S/m\lesssim r_0\ll M \,.
\eean
One sees immediately that the turning point given in \eq{rs} cannot exist outside the black hole.

Now the only possibility for the divergence is that $\Delta $ in the second denominator of \eq{eef} vanishes,
 i.e., the collision must occur at the outer horizon $r=r_+$. To check if $E_{eff}$ could be infinite,we expand $E^\prime$ at $r=r_+$ and find
\be
E^\prime=\alpha_0+\alpha_1(r-r_+)+\alpha_2(r-r_+)^2+ \dots, \label{alpha1}
\ee
where
\be
\nn
\alpha_0&=&-r_+^2[a^2Er_++Er_+^3-Js+a(-Jr_++Es+Er_+s)]\\\nn
&&[a^2E_1r_++E_1r_+^3-J_1s_1+a(-J_1r_++E_1s_1+E_1r_+s_1)]\\
&+&\sqrt{O\mid_{r=r_+}}\sqrt{O_1\mid_{r=r_+}} \,.
\ee
At the horizon, \eq{bigo} takes the form
\bean
O\mid_{r=r_+}=r_+^2\left[J(a r_++s)-E(a^2 r_++r_+^3+as+a r_+s)\right]^2  \label{or}\,.
\eean
Using \eq{jin}, the square root terms can be simplified, and one finds
\be
\alpha_0=0 \,.
\ee
The vanishing of $\alpha_0$ is important because it rules out the divergence of $E_{c.m.}$ for generic angular momentums.

The second term on the right-hand side of \eq{eef} is essential for the divergence of $E_{eff}$, which is proportional to $\frac{E'}{\Delta}$. Note that near the horizon, $\Delta\sim (r-r_+)^2$ for the extremal case $a=1$, and $\Delta\sim r-r_+$ for the non-extremal case $a<1$. Therefore, we need to discuss them separately.

\subsection{Extremal case $a=1$}
Now the horizon is located at $r=1$. It is straightforward to check, from \eq{ep19}, that
\bean
\alpha_1=\frac{dE_p}{dr}\Big|_{r=1}=0 \,.
\eean
Therefore, the second term in \eq{eef} takes the form
\bean
\frac{E'}{\Delta}\sim \frac{\alpha_2 (r-1)^2}{\Delta}\sim \alpha_2 \,.
\eean
The expression of $\alpha_2$ is found to be
\bean
\alpha_2=\frac{d^2E_p}{dr^2}\Big|_{r=1}=\frac{\beta}{(J-2E)(J_1-2E_1)(s+1)(s_1+1)}
\eean
The expression of the nominator $\beta$ is rather lengthy. But we see from the denominator that when
\bean
J=J_c=2E \label{jce} \,,
\eean
$\alpha_2$ will be divergent. Consequently, $E_{eff}$ will blow up. Note that the critical angular momentum in \eq{jce} is the same as that when the spin is absent \cite{bsw}, but both $E$ and $J$ have been redefined with the spin correction.

We also need to make sure that the particle can actually reach the horizon, i.e., \eq{17} should be satisfied. When $M=a=1$ and $J=2E$, it is straightforward to see that
\bean
O\rightarrow(r-1)^2 \left[E^2 r^2 \left(r^4+2 r^3-r^2 (s-4) s+2 r s+s^2\right)-\left(r^3-s^2\right)^2\right]
\eean
as $r\rightarrow 1$. Since $E=\tilde E/m>1$ and $s$ is small, we see immediately that
\bean
O>0
\eean
for $r\sim 1$. Therefore, the particle with critical angular momentum can reach the horizon.

Finally, we need to check \eq{pb}, i.e., $v^a$ is timelike. For $M=a=r=1$ and $J=2E$, \eq{pp} reduces to
\bean
P=-3 E^2 (s-1)^2 s^4-6 E^2 (s-1)^2 s^2+s^7-4 s^6+6 s^4-4 s^2+1\,.
\eean
Since $s\ll1$, we see $P>0$ near the horizon. Thus, \eq{pb} is verified.

\subsection{Non-extremal case $0<a<1$}
Now \eq{eef} suggests that $E_{eff}$ can be infinite if $\alpha_1$ is infinite. $\alpha_1$ can be obtained by taking derivative of $E'$ in \eq{ep19}. Obviously, $\sqrt{O\mid_{r=r_+}}$ appears in the denominator of $\alpha_1$. Hence, an infinite $\alpha_1$ requires
\be
O\mid_{r=r_+}=0 \,,
\ee
which means, according to \eq{or}, that the angular momentum must take the critical value
\be
J=J_c=\frac{E \left[\left(\sqrt{1-a^2}+2\right) a s-2 a^2+4 \left(\sqrt{1-a^2}+1\right)\right]}{ar_++s}  \,,\label{jjc}
\ee
However,to make sure that the particle with this critical angular momentum can actually reach the horizon, Eq.(\ref{17}) must hold outside the horizon.

 By Taylor expansion, we find
\be\label{21}
O=a_0+a_1(r-r_+)+\dots \,,
\ee
For $J=J_c$, one can show $a_0=0$. After some highly non-trivial algebra manipulation, we find that $a_1$ can be written in the form
\be
a_1=b_1+E^2b_2\,,
\ee
where
\bean
b_1\eqn -2 \sqrt{1-a^2} \left[\left(\sqrt{1-a^2}+3\right) a^2-4 \left(\sqrt{1-a^2}+1\right)+s^2\right]^2 \,,\\ \non
b_2&=&-2\frac{8(1+\sqrt{1-a^2})+a^4(4+\sqrt{1-a^2})-4a^2(3+2\sqrt{1-a^2})}{(a^4+2as+s^2)^2}\non
&&\left[\left(\sqrt{1-a^2}+3\right) a^2-4 \left(\sqrt{1-a^2}+1\right)+s^2\right]^2
\left[s-a(-1+\sqrt{1-a^2})\right]^2 \,.\non
\
\eean
Simple analysis shows that both $b_1$ and $b_2$ are non-positive for $0<a<1$. It is also easy to see that $b_1$ and $b_2$ cannot be zero because $s$ is small. Therefore
\be
a_1<0\,,
\ee
which means $O$ is negative near the horizon and consequently the spinning particle with $J=J_c$ cannot approach the horizon.

\section{Conclusions}
We have shown that arbitrarily high energy can be obtained near the horizon of an extremal black hole if one particle with spin possesses the critical angular momentum. Although the definition of conserved angular momentum and energy have been modified due to the particle's spin, the critical angular momentum $J=2E$ remains unchanged. So this relation may be generalized from point particles to spinning particles. For non-extremal black holes, we show that unlimited collision energy can not be found even if the particle's spin is taken into account. Obviously, a spinning particle is closer to a real particle than a point particle which follows geodesic motions. So our work further extends previous studies on particle collisions near black holes.

\section*{Acknowledgements}
 This research was supported by NSFC Grants No. 11235003, No. 11375026 and NCET-12-0054.

\end{document}